%% file: main.tex
\documentclass[letterpaper]{article} 
\usepackage{aaai24}  
\usepackage{times}  
\usepackage{helvet}  
\usepackage{courier}  
\usepackage[hyphens]{url}  
\usepackage{graphicx} 
\def\UrlFont{\rm}  
\usepackage{natbib}  
\usepackage{caption} 
\usepackage{algorithm}
\usepackage{algorithmic}

\usepackage{newfloat}
\usepackage{listings}
\DeclareCaptionStyle{ruled}{labelfont=normalfont,labelsep=colon,strut=off} 
\floatstyle{ruled}
\newfloat{listing}{tb}{lst}{}
\floatname{listing}{Listing}
\title{A Systematic Study of Joint Representation Learning \\ on Protein Sequences and Structures}
\author {
    Zuobai Zhang\textsuperscript{\rm 1,2},
    Chuanrui Wang\textsuperscript{\rm *,1,2},
    Minghao Xu\textsuperscript{\rm *,1,2}, \\
    Vijil Chenthamarakshan\textsuperscript{\rm 3},
    Aur\'{e}lie Lozano\textsuperscript{\rm 3},
    Payel Das\textsuperscript{\rm 3},
    Jian Tang\textsuperscript{\rm 1,4,5}
}
\affiliations {
    \textsuperscript{\rm *} equal contribution
    \textsuperscript{\rm 1} Mila - Qu\'ebec AI Institute
    \textsuperscript{\rm 2} Universit\'e de Montr\'eal\\
    \textsuperscript{\rm 3} IBM Research
    \textsuperscript{\rm 4} HEC Montr\'eal
    \textsuperscript{\rm 5} CIFAR AI Chair\\
    \texttt{\{zuobai.zhang,chuanrui.wang,minghao.xu\}@mila.quebec}, \\
    \texttt{\{ecvijil,aclozano,daspa\}@us.ibm.com},
    \texttt{jian.tang@hec.ca}
}

\usepackage{bibentry}

\input{command}
\input{math}

\begin{document}

\maketitle

\input{sections/00_abstract}

\input{sections/01_introduction}

\input{sections/02_related}

\input{sections/03_pretrain}
\input{sections/04_experiment}

\input{sections/05_conclusion}

\section*{Acknowledgments}

This project is supported by AIHN IBM-MILA partnership program, the Natural Sciences and Engineering Research Council (NSERC) Discovery Grant, the Canada CIFAR AI Chair Program, collaboration grants between Microsoft Research and Mila, Samsung Electronics Co., Ltd., Amazon Faculty Research Award, Tencent AI Lab Rhino-Bird Gift Fund, a NRC Collaborative R\&D Project (AI4D-CORE-06) as well as the IVADO Fundamental Research Project grant PRF-2019-3583139727.

\bibliography{references}



\end{document}

%% file: command.tex
\usepackage{booktabs}       

\usepackage{enumitem}
\usepackage{xspace}
\usepackage{xcolor}
\usepackage{caption}
\usepackage{amssymb}
\usepackage{multirow}
\usepackage{adjustbox}
\usepackage{subfigure}
\usepackage{makecell}
\usepackage{pifont}
\usepackage{bbding}
\usepackage{threeparttable}
\usepackage{babel}

\usepackage{thmtools}
\usepackage{thm-restate}

\newcommand{\method}{GearNet\xspace}
\newcommand{\algo}{MMPreT\xspace}

\newcommand{\mathbbm}[1]{\text{\usefont{U}{bbm}{m}{n}#1}}

\definecolor{myblue}{RGB}{68, 114, 196}
\definecolor{myorange}{RGB}{237, 125, 49}

\setlist[enumerate]{leftmargin=7.5mm}

%% file: math.tex
\usepackage{amsmath}
\usepackage{bm}

\def\eqref#1{equation~\ref{#1}}

\def\1{\bm{1}}

\def\ve{{\bm{e}}}

\def\vh{{\bm{h}}}

\def\vu{{\bm{u}}}

\def\vx{{\bm{x}}}

\def\vz{{\bm{z}}}

\def\mW{{\bm{W}}}

\DeclareMathAlphabet{\mathsfit}{\encodingdefault}{\sfdefault}{m}{sl}
\SetMathAlphabet{\mathsfit}{bold}{\encodingdefault}{\sfdefault}{bx}{n}

\def\gE{{\mathcal{E}}}

\def\gG{{\mathcal{G}}}

\def\gL{{\mathcal{L}}}

\def\gN{{\mathcal{N}}}

\def\gP{{\mathcal{P}}}

\def\gR{{\mathcal{R}}}

\def\gV{{\mathcal{V}}}

\def\gX{{\mathcal{X}}}

\newcommand{\E}{\mathbb{E}}

\newcommand{\R}{\mathbb{R}}

\let\oldAA\AA
\renewcommand{\AA}{\text{\normalfont\oldAA}}
\usepackage{cancel}

%% file: sections/00_abstract.tex

\begin{abstract}
Learning effective protein representations is critical in a variety of tasks in biology such as predicting protein functions.
Recent sequence representation learning methods based on Protein Language Models (PLMs) excel in sequence-based tasks, but their direct adaptation to tasks involving protein structures remains a challenge.
In contrast, structure-based methods leverage 3D structural information with graph neural networks and geometric pre-training methods show potential in function prediction tasks, but still suffers from the limited number of available structures.
To bridge this gap, our study undertakes a comprehensive exploration of joint protein representation learning by integrating a state-of-the-art PLM (ESM-2) with distinct structure encoders (GVP, GearNet, CDConv). 
We introduce three representation fusion strategies and explore different pre-training techniques. 
Our method achieves significant improvements over existing sequence- and structure-based methods, setting new state-of-the-art for function annotation. 
This study underscores several important design choices for fusing protein sequence and structure information.
Our implementation is available at \url{https://github.com/DeepGraphLearning/ESM-GearNet}.
\end{abstract}


%% file: sections/01_introduction.tex

\section{Introduction} \label{sec:intro}


Proteins, as fundamental building blocks of life, play a pivotal role in numerous biological processes, ranging from enzymatic reactions to cellular signaling. 
Their intricate three-dimensional structures and dynamic behaviors underscore their functional diversity. 
Effective understanding of proteins is crucial for unraveling mechanisms underlying diseases, drug discovery, and synthetic biology. 
Herein, protein representation learning has emerged as a highly promising avenue, showcasing its efficacy across diverse protein comprehension tasks, such as protein structure prediction~\citep{jumper2021highly,baek2021accurate}, protein function annotation~\citep{gligorijevic2021structure,meier2021language,zhang2022protein}, protein-protein docking~\citep{corso2023diffdock,zhang2023ebind} and protein design~\citep{hsu2022learning,dauparas2022robust}.

Given the recent strides in the advancement of large pre-trained language models for natural languages~\citep{vaswani2017attention,devlin2018bert,NEURIPS2020_1457c0d6}, various categories of language models  have been repurposed for protein representation learning. 
These protein language models (PLMs) consider protein sequences as the essence of life's language, treating individual amino acids as tokens.
Self-supervised learning methods are applied to acquire informative protein representations from billions of natural protein sequences.
Notable instances include long short-term memory (LSTM)-based PLMs like UniRep~\citep{alley2019unified}, as well as transformer-based PLMs like ProtTrans~\citep{elnaggar2020prottrans}, Ankh~\citep{elnaggar2023ankh} and ESM~\citep{rives2021biological,lin2023evolutionary}.
While these methods exhibit substantial potential in protein function prediction tasks~\citep{tape2019,xu2022peer}, their direct application to tasks involving structural inputs, such as protein structure assessment and protein-protein interaction prediction, presents challenges.

Inspired by advancements in protein structure prediction tools~\citep{jumper2021highly,lin2023evolutionary} and the critical role of protein structures in determining functionality, another strand of methods focuses on acquiring protein representations based on 3D structures.
These approaches model proteins as graphs, with atoms or amino acids serving as nodes and edges indicating spatial adjacency.
Subsequently, 3D graph neural networks (GNNs) facilitate message propagation to capture interactions between residues, enabling the extraction of representations invariant to structural translation and rotation. 
Typical examples include GearNet~\citep{zhang2022protein}, GVP~\citep{jing2021equivariant},  CDConv~\citep{fan2023continuousdiscrete}.
Additionally, efforts have been made to design pre-training strategies that leverage unlabeled protein structures from PDB~\citep{berman2000protein} and the AlphaFold Database~\citep{varadi2021alphafold}.
These methods rely on self-supervised learning techniques such as contrastive learning~\citep{zhang2022protein,chen2022structure}, self-prediction~\citep{zhang2022protein}, and denoising~\citep{Guo2022SelfSupervisedPF,zhang2023siamdiff}, enabling structure encoders to achieve top-tier performance on tasks related to protein structure, even pre-trained on a relatively small set of unlabeled proteins.
Nonetheless, these structure-based approaches still suffer from the limited number of available structures compared with PLMs, raising questions about their ability to surpass sequence-based methods.

In order to understand how to combine the advantages of both worlds, we conduct a comprehensive investigation into joint protein representation learning.
Our study combines a state-of-the-art PLM (ESM-2) with three distinct structure encoders (GVP, GearNet, and CDConv). 
We introduce three fusion strategies—serial, parallel, and cross fusion—to combine sequence and structure representations.
We further explore six diverse pre-training techniques, employing the optimal model from the aforementioned choices and leveraging pre-training on the AlphaFold Database.
Our findings indicate that:
\begin{enumerate}
    \item Serial fusion, a straightforward approach, proves remarkably effective, outperforming the other two fusion strategies across most tasks.
    \item Adapting a reduced learning rate for PLMs is crucial to safeguard their representations from disruption.
    \item Despite GearNet's relative performance lag behind the other encoders, it demonstrates superior results after integration with PLMs.
    \item The two pre-training methods leveraging both sequence and structure information can yield superior performance compared to other methods relying solely on either sequence or structure information.
\end{enumerate}
Drawing from these insights, our method achieves significant improvements over existing sequence- and structure-based methods, establishing a new state-of-the-art on Enzyme Commission and Gene Ontology annotation tasks.
We believe that this work holds practical significance in the adaptation of PLMs with structure-based encoders.

%% file: sections/02_related.tex
\section{Related Work} \label{sec:rela}

\textbf{Sequence-based representation learning.}
Regarding protein sequences as the language of life, models from the rapidly developing field of NLP are widely used in modeling protein sequence data. Examples include the CNN-based models~\citep{shanehsazzadeh2020transfer}, LSTM-based models~\citep{tape2019}, ResNet~\cite{tape2019} and transformer-based models~\citep{elnaggar2020prottrans,rives2021biological,lin2023evolutionary,zhang2022ontoprotein,xu2023protst,notin2022tranception,madani2023large,chen2023xtrimopglm}. 
Given the rising number of protein sequences and the substantial cost of labeling their functions, representation learning is typically conducted in a self-supervised manner to leverage the extensive protein sequence datasets, via autoregressive modeling~\citep{notin2022tranception,madani2023large,hesslow2022rita,elnaggar2021prottrans,elnaggar2023ankh}, masked language modeling (MLM)~\citep{elnaggar2020prottrans,rives2021biological,lin2023evolutionary}, pairwise MLM~\citep{he2021pre}, contrastive learning~\citep{lu2020self}, \emph{etc.} PLMs have shown impressive performance on capturing underlying patterns of sequences, thus predicting protein structures~\citep{lin2023evolutionary} and functionality~\citep{tape2019,xu2022peer, chen2023xtrimopglm}. 
However, these existing PLMs cannot explicitly encode  protein structures, which are actually determinants of diverse protein functions. In this work, we seek to overcome this limitation by enhancing a PLM with a protein structure encoder so as to capture detailed protein structural characteristics. 



\textbf{Structure-based representation learning.} Diverse types of protein structure encoders have been devised to capture different granularities of protein structures, including residue-level structures~\citep{gligorijevic2021structure,zhang2022protein,xu2023eurnet, jing2021equivariant,hsu2022learning, dauparas2022robust}, atom-level structures~\citep{jing2021equivariant,hermosilla2020intrinsic} and protein surfaces~\citep{gainza2020deciphering,sverrisson2021fast}. These structure encoders have boosted protein function understanding~\citep{gligorijevic2021structure,zhang2022protein}, protein design~\citep{jing2021equivariant,hsu2022learning, dauparas2022robust, gao2023pifold} and protein structure generation~\citep{wu2022protein, trippe2023diffusion}.
Various self-supervised learning algorithms are designed to learn informative protein structure representations, including contrastive learning~\citep{zhang2022protein,hermosilla2022contrastive}, self-prediction~\citep{zhang2022protein,chen2022structure}, denoising score matching~\citep{Guo2022SelfSupervisedPF,wu2022pre} and structure-sequence multimodal diffusion~\citep{zhang2023siamdiff}. Structurally pre-trained models outperform PLMs on function prediction tasks~\citep{zhang2022ontoprotein,hermosilla2022contrastive}, given the principle that protein structures are the determinants of their functions. 

\textbf{Joint representation learning.} 
The integration of protein sequence-based models with protein structure models remains unexplored. 
Early attempts like LM-GVP sought to combine PLMs with structure encoders~\citep{wang2022lm}. 
While recent methods have been introduced, their outcomes have not consistently surpassed those of single-modality models~\citep{huang2023data, heinzinger2023prostt5}. 
In this study, we introduce three novel fusion methods aimed at harnessing the bimodal information of both sequence and structure.
Different from earlier approaches, we emphasize the potential benefits of leveraging bimodal data. 
This is achieved by incorporating sequential information into distinct residue-level encoders—GearNet, GVP, and CDConv\citep{zhang2022protein, jing2021equivariant, fan2023continuousdiscrete}. 
Furthermore, we enhance the effectiveness of the proposed sequence-structure hybrid encoder, ESM-GearNet, through structure-based pre-training.

%% file: sections/03_pretrain.tex
\section{Methods}

In this section, we describe basic concepts of proteins and sequence- and structure-based protein representation learning methods.
Next, we propose three different strategies for combining sequence and structure representations. 
Finally, we present how different pre-training algorithms can be applied on the proposed architecture.

\subsection{Proteins}
Proteins are large molecules composed of residues, \emph{a.k.a.} amino acids, linked together in chains. 
Despite there being only 20 standard residue types, their numerous combinations contribute to the immense diversity of proteins found in nature. 
The specific arrangement of these residues determines the 3D positions of all the atoms within the protein, forming what we call the protein's structure.
A residue includes elements like an amino group, a carboxylic acid group, and a side chain group that defines its type. 
These components connect to a central carbon atom known as the alpha carbon.
For simplicity in our work, we use only the alpha carbon atoms to represent the main backbone structure of each protein.
Each protein can be represented as a sequence-structure pair $\gP=(\gR,\gX)$, where  $\gR=[r_1,r_2,\cdots,r_{n}]$ denotes the sequence of the protein with $r_i\in\{1,...,20\}$ indicating the type of the $i$-th residue, and $\gX=[\vx_1,\vx_2...,\vx_{n}]\in\R^{n\times 3}$ denotes its structure with $\vx_i$ representing the Cartesian coordinates of the $i$-th alpha carbon atom, and $n$ denotes the number of residues.

\subsection{Sequence-Based Protein Representation Learning}

Treating protein sequences as the language of life, recent works draw inspirations from large pre-trained language models to learn the evolutionary information from billions of protein sequences via self-supervised learning.
In this work, we illustrate our method using the transformer-based Protein Language Model (PLM) ESM~\citep{rives2021biological,lin2023evolutionary}.
These models process residue type sequences through multiple self-attention layers and feed-forward networks to capture inter-residue dependencies. 
Specifically, we represent the hidden state of the $i$-th residue as $\vh^{(l)}_i$, initialized with the residue's type embedding $\vh^{(0)}_i = \text{Embedding}(r_i)\in\R^{d}$, where $d$ denotes the hidden representation dimension.
Self-attention layers compute attention coefficients $\alpha_{ij}$, measuring residue contact strength between $i$ and $j$. 
The output representations are further fed into forward networks.
\begin{align*}
    &\alpha^{(l)}_{ij} = \begin{matrix}\text{Softmax}_j(\frac{1}{\sqrt{d}}\text{Linear}_q(\vh^{(l)}_i)\cdot \text{Linear}_k(\vh^{(l)}_j))\end{matrix}\\
    &\vh^{(l+0.5)}_{i} = \begin{matrix}\vh^{(l)}_{i} + \sum_j \alpha^{(l)}_{ij}\cdot \text{Linear}_v(\vh^{(l)}_j)\end{matrix}\\
    &\vh^{(l+1)}_{i} = \vh^{(l+0.5)}_{i} + \text{FeedForward}(\vh^{(l+0.5)}_{i})
\end{align*}
In practice, positional embeddings, multi-head attention and layer norm layers are incorporated, enhancing the modeling process (details omitted here).

These models are pre-trained with masked language modeling (MLM) loss by predicting the type of a masked residue given the surrounding context.
An additional linear head employs the final-layer representations $\vh^{(L)}$ for the prediction.
The loss function for each sequence is
\begin{align*}
    \gL_{MLM} = \begin{matrix}\E_M [\sum\nolimits_{i\in M} - \log p(r_i|r_{/M})]\end{matrix},
\end{align*}
where a random set of indices $M$ is chosen for masking, replacing the true token at each index $i$ with a mask token.
For each masked token, the loss aims to minimize the negative log likelihood of the true residue $r_i$ given the masked sequence $r_{/M}$ as context.
By fully utilizing massive unlabeled data, these models have achieved state-of-the-art performance on various protein understanding tasks~\citep{lin2023evolutionary,Elnaggar2023AnkhO}.

\subsection{Structure-Based Protein Representation Learning}

\begin{figure*}[t]
    \centering
    \includegraphics[width=\linewidth]{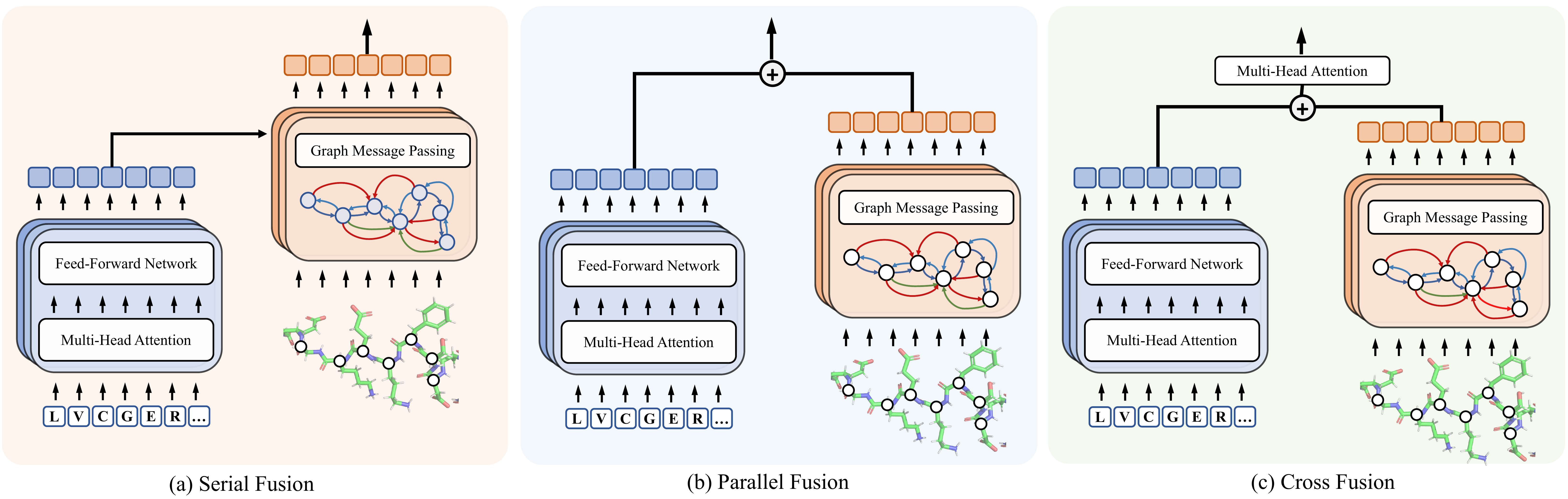}
    \caption{Three different ways to fuse sequence and structure representations. (a) \emph{Serial fusion}, where sequence representations are used as residue features in structure encoders.
    (b) \emph{Parallel fusion}, involving the concatenation of sequence and structure representations. (c) \emph{Cross fusion}, where sequence and structure representations are combined via multi-head self-attention.}
    \label{fig:esm_gearnet}
\end{figure*}

The achievements of AlphaFold2~\citep{jumper2021highly} have revolutionized precise protein structure prediction, triggering a wave of research on structure-driven pre-training~\citep{zhang2022protein,chen2022structure,zhang2023siamdiff} due to the direct influence of structures on protein functionalities.
Given a protein, structure-based techniques often establish a graph incorporating both sequential and spatial details, leveraging graph neural networks to learn representations.
In this work, we focus on three commonly used protein structure  encoder: GearNet, GVP and CDConv.

\paragraph{GearNet~\citep{zhang2022protein}}
GearNet represents proteins using a multi-relational residue graph $\gG=(\gV,\gE,\gR)$, where $\gV$ and $\gE$ denote the sets of residues and edges, respectively, and $\gR$ represents the edge types. 
Three directed edge types are incorporated into the graph: sequential edges, radius edges, and K-nn edges. Specifically,
\begin{align*}
    \gE^{(\text{seq})} &= \{(i,j)|i,j\in\gV,|j-i|<d_{\text{seq}}\}, \\
    \gE^{(\text{radius})} &= \{(i,j)|i,j\in\gV,|\vx_j-\vx_i|<d_{\text{radius}}\}, \\
    \gE^{(\text{knn})} &= \{(i,j)|i,j\in\gV,j\in \text{knn}(i)\}, \\
    \gE &= \gE^{(\text{seq})} \cup \gE^{(\text{radius})} \cup \gE^{(\text{knn})},
\end{align*}
where $d_{\text{seq}}=3$ defines the sequential distance threshold, $d_{\text{radius}}=10\mathrm{\AA}$ defines the spatial distance threshold, and $\text{knn}(i)$ indicates the K-nearest neighbors of node $i$ with $k=10$.
For sequential edges, edges with different sequential distances are treated as different types.
These edge types collectively reflect distinct geometric attributes, contributing to a holistic featurization of proteins.
Upon constructing the graph, a relational message passing procedure is conducted.
We denote $\vu^{(l)}$ as the representations at the $l$-th layer, initialized with $\vu^{(0)}_i = \text{Embedding}(r_i)$.
The message passing process can be written as:
\begin{align*}
    \vu^{(l)}_i &= \begin{matrix}
        \vu^{(l-1)}_i +\sigma\left(\sum\nolimits_{r\in\gR}\mW_r\sum\nolimits_{j\in \gN_r(i)} \vu^{(l-1)}_j\right)
    \end{matrix},
\end{align*}
where $\gN_r(i)$ is the set of neighbors of $i$ with edge type $r$, and $\sigma(\cdot)$ is the ReLU function. 

\paragraph{GVP~\citep{jing2021equivariant}}
The GVP module replaces standard MLPs in GNN aggregation and feed-forward layers, operating on scalar and geometric features—features that transform as vectors with spatial coordinate rotations
A radius graph is constructed with $\gE=\gE^{(\text{radius})}$ where $d_{\text{radius}}=10\mathrm{\AA}$.
A radius graph, $\gE=\gE^{(\text{radius})}$ with $d_{\text{radius}}=10\mathrm{\AA}$, is constructed. 
Node features begin as $\vu^{(0)}_i = (\text{Embedding}(r_i), \mathbf{0})$, while edge features are $\ve{(j,i)} = (\text{rbf}(\vx_j-\vx_i), \vx_j-\vx_i)$, using $\text{rbf}(\cdot)$ for pairwise distance features.
For message functions, the GVP network concatenates node and edge features, applying the GVP module for message passing on the (scalar, vector) representations. 
A feed-forward network follows each message passing layer.
Formally,
\begin{align*}
    &\vu^{(l+0.5)}_{i} =\begin{matrix}
        \vu^{(l)}_{i}+\frac{1}{|\gN(i)|}\sum_{j\in\gN(i)}\text{GVP}([\vu^{(l)}_{j}, \ve_{(j,i)}])
    \end{matrix},\\
    &\vu^{(l+1)}_{i} = \begin{matrix}
        \vu^{(l+0.5)}_{i}+\text{GVP}(\vu^{(l+0.5)}_{i})
    \end{matrix},
\end{align*}
where $\vu_i^{(l)} \in\R^d\times\R^{d'\times 3}$ denotes hidden representation tuples at the $l$-th layer, and $\gN(i)$ is neighbors of $i$.
$\text{GVP}(\cdot)$ is the proposed module maintaining SE(3)-invariance of scalar features and SE(3)-equivariance of vector features.
The scalar features at the last layer of each node are utilized for property prediction to keep SE(3)-invariance.

\paragraph{CDConv~\citep{fan2023continuousdiscrete}}
CDConv adopts GearNet's concept of multi-type message passing to capture sequential and spatial interactions among residues.
Instead of using distinct kernel matrices for varied edge types, CDConv employs an MLP to parameterize kernel matrices, relying on relative spatial and sequential information between two residues.
With the same initialization as GearNet, the message passing procedure is written as
\begin{align*}
    \vu^{(l)}_i &= \begin{matrix}
        \vu^{(l-1)}_i +\sigma\left(\sum\nolimits_{j\in \gN(i)}\mW(\vx_j-\vx_i,j-i)\;\vu^{(l-1)}_j\right)
    \end{matrix},
\end{align*}
where $\mW(\cdot,\cdot)$ represents an MLP that takes relative positions in Euclidean space and sequences as input, producing the kernel matrix as output.
The edge set is the intersection of sequential and spatial edges $\gE = \gE^{(\text{seq})} \cap \gE^{(\text{radius})}$ with $d_{\text{seq}}=11$.
To reduce node counts and expand reception field, a half pooling approach is employed every two CDConv layers, merging adjacent nodes.
For the $i$-th CDConv layer, the radius is set to $\lceil i/2+1\rceil d_{\text{radius}}$, and the output dimension is set to $\lceil i/2+1 \rceil d$, where $d_{\text{radius}}=4\mathrm{\AA}$ and $d$ denote the initial radius and initial hidden dimensions, respectively.
Due to the pooling scheme, CDConv cannot yield residue-level representations.

\subsection{Fusing Sequence and Structure Representations}
\label{sec:fusion}

While protein language models implicitly capture structural contact information, explicitly incorporating detailed structures can effectively model spatial interactions among residues.
Huge-scale pre-training of PLMs also significantly bolsters relatively small protein structure encoders. 
In this subsection, we propose fusing representations from protein language models and protein structure encoders, presenting three fusion strategies illustrated in Figure~\ref{fig:esm_gearnet}:
\begin{enumerate}[leftmargin=1.3em]
\item \emph{Serial fusion}. Rather than initializing structure encoder input node features with residue type embeddings, we initialize them with PLM outputs, denoted as $\vu^{(0)}=\vh^{(L)}$, and utilize the structure encoder's output as the final protein representations, $\vz=\vu^{L}$. 
This approach provides more powerful residue type representations incorporating sequential context.

\item \emph{Parallel fusion}. 
We concatenate outputs of sequence encoders and structure encoders for final representations, yielding $\vz=[\vh^{(L)},\vu^{(L)}]$. 
This fusion method combines both representations while keeping the structure encoder from affecting pre-trained sequence representations.

\item \emph{Cross fusion}. 
To enhance interaction, we introduce a cross-attention layer over sequence and structure representations as $\vz_i=\text{SelfAttn}([\vh_i^{(L)},\vu_i^{(L)}])$. The attention layer's output is averaged over the protein to produce final representations $\vz$. 
\end{enumerate}
Ultimately, the resulting representation is employed for residue-level or protein-level predictions.

\paragraph{Reduced learning rate of PLMs}
Given that structure encoders start from random initialization while PLMs are pre-trained, we've observed practical benefits in utilizing a lower learning rate for PLMs to prevent  catastrophic forgetting.
In our experiments, we maintain a learning rate ratio of $0.1$, a strategy we find crucial for the robust generalization of our proposed fusion approaches.

\begin{figure*}[t]
    \centering    
    \includegraphics[width=\linewidth]{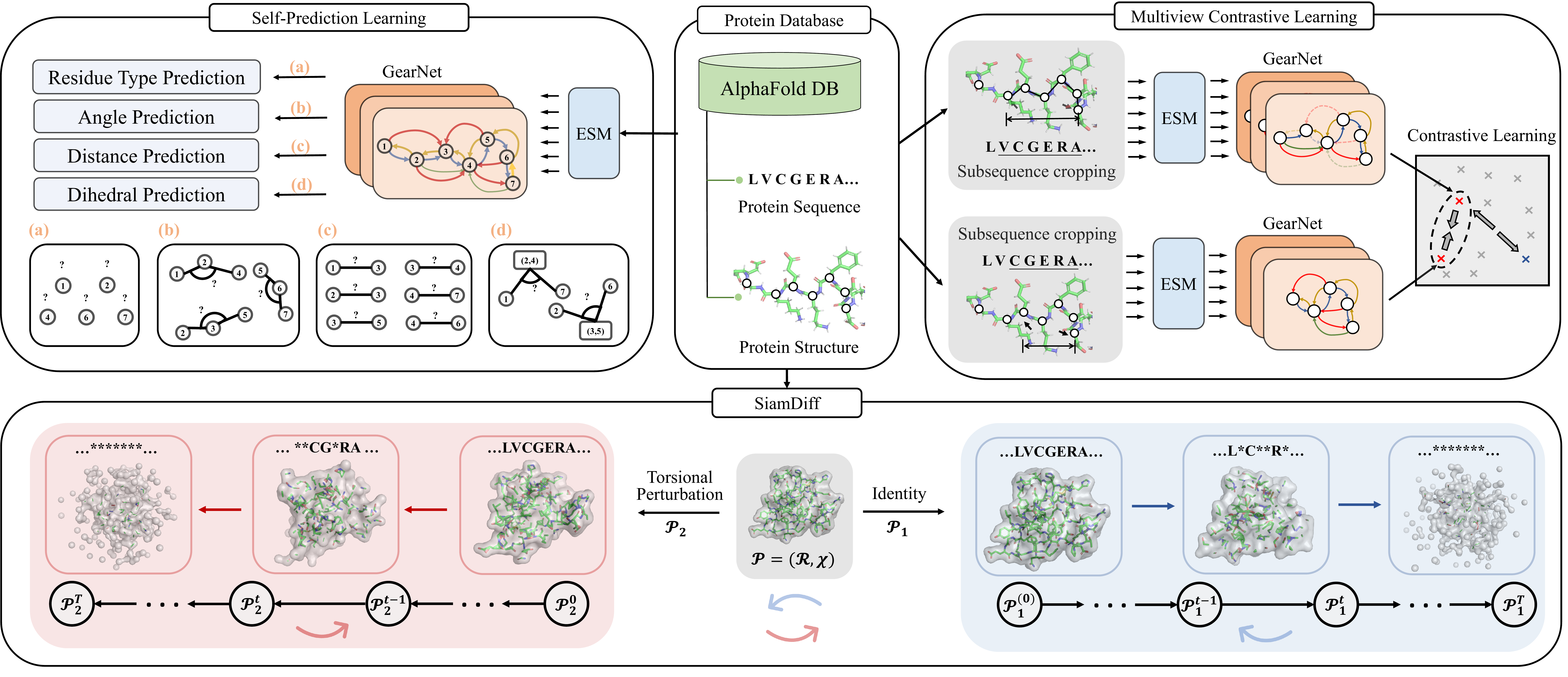}
    \caption{
    Pre-training ESM-GearNet on AlphaFold Database with six different methods: residue type prediction, angle prediction, distance prediction, dihedral prediction, multiview contrastive learning and SiamDiff.
    }
    \label{fig:esm_gearnet_mc}
\end{figure*}

\subsection{Joint Pre-Training on Unlabeled Proteins}

The current joint encoder effectively utilizes knowledge acquired from extensive unlabeled protein sequences.
Recent strides in accurate protein structure prediction, have provided access to a substantial collection of precise protein structures, such as AlphaFold Database~\citep{varadi2021alphafold}.
Several structure-based pre-training techniques have emerged, including self-prediction~\citep{zhang2022protein}, multiview contrast~\citep{chen2022structure}, and denoising objectives~\citep{zhang2023siamdiff}. 
Taking a step further, we next discuss how to apply these pre-training algorithms upon the joint encoder, as illustrated in Figure~\ref{fig:esm_gearnet_mc}.
During pre-training, ESM remains fixed while only the structure encoder is tuned, preserving sequence representations.

\textbf{Self-prediction methods~\citep{zhang2022protein}.}
Based on the recent progress of self-prediction methods in natural language processing~\citep{devlin2018bert,NEURIPS2020_1457c0d6}, these methods aim to predict one part of the protein given the remaining context.
Four self-supervised tasks are introduced, guided by  geometric attributes.
These methods perform masked prediction on individual residues, residue pairs, triplets, and quadruples, subsequently predicting residue types, distances, angles, and dihedrals, respectively.
The corresponding loss functions are summarized in Table~\ref{tab:self_prediction}.
\begin{table}[t]
    \centering
    \begin{adjustbox}{max width=\linewidth}
        \begin{tabular}{lc}
            \toprule
            \bf{Method}  & \bf{Loss function} \\
            \midrule
            \bf{Residue Type Prediction}   & $ \text{CE}(f_{\text{res}}(\vz_i), r_i)$ \\
            \bf{Distance Prediction}   & $(f_{\text{dis}}(\vz_i,\vz_j) - \|\vx_i-\vx_j\|_2)^2$\\
            \bf{Angle Prediction}   & $\text{CE}(f_{\text{ang}}(\vz_i,\vz_j,\vz_k), \text{bin}(\angle ijk))$ \\
            \bf{Dihedral Prediction}   & $ \text{CE}(f_{\text{dih}}(\vz_i,\vz_j,\vz_k,\vz_t), \text{bin}(\angle ijkt))$ \\
            \bottomrule
        \end{tabular}
    \end{adjustbox}
    \caption{Self-prediction methods. We use $i$,$j$,$k$,$t$ to denote sampled residue indices.
    Tasks are associated with respective MLP heads: $f_{\text{res}}$, $f_{\text{dis}}$, $f_{\text{ang}}$, and $f_{\text{dih}}$.
    $\text{CE}(\cdot)$ is the cross entropy loss, and angles are discretized with $\text{bin}(\cdot)$. }
    \label{tab:self_prediction}
\end{table}

\textbf{Multiview contrastive learning~\citep{zhang2022protein}.}
The frameworks aim to maintain similarity between correlated protein subcomponents after mapping to a lower-dimensional latent space.
For a protein graph $\gG$, we utilize \emph{subsequence cropping} to randomly select consecutive subsequences.
This scheme captures protein domains—recurring consecutive subsequences in different proteins that signify functions~\citep{Ponting2002TheNH}.
After subsequence sampling, following common self-supervised learning practice~\citep{chen2020simple}, we employ a noise function for diverse views, specifically \emph{random edge masking} that hides $15\%$ of edges in the protein graph.
We align their representations in the latent space with an InfoNCE loss~\citep{chen2020simple}.
Let $x,y$ represent subcomponent graphs from the same protein, and $k$ from other proteins within the same batch, with corresponding representations $\vz_x,\vz_y,\vz_k$.
The loss function is written as
\begin{align*}
    \gL_{x,y} = - \log \frac{\exp(\text{sim}(g(\vz_x),g(\vz_y))/\tau)}{\sum_{k=1}^{2B}\mathbbm{1}_{[k\neq x]}\exp(\text{sim}(g(\vz_y),g(\vz_k))/\tau)},
\end{align*}
where $g(\cdot)$ denotes an MLP applied to latent representations, $B,\tau$ denote batch size and temperature, and $\mathbbm{1}{[k\neq x]}\in\{0,1\}$ acts as an indicator function that equals $1$ \emph{iff} $k\neq x$.

\textbf{Diffusion-based pre-training~\citep{zhang2023siamdiff}.}
These methods are inspired by the success of diffusion models in capturing the joint distribution of sequences and structures.
During pre-training, noise levels $t\in\{1,..,T\}$ are sampled and applied to structures and sequences, where higher levels indicate larger noise.
The encoder's representations are used for denoising with loss functions:
\begin{align*}
    \gL_{\text{struct}}
    &=\begin{matrix}\E_{t\sim \{1,..,T\}} \E_{\epsilon\sim \gN(0,I)}\left[\|\epsilon - f_{\text{noise}}(\vz^{(t)},\vx^{(t)})\|_2^2\right]\end{matrix},\\
    \gL_{\text{seq}}
    &=\begin{matrix}\E_{t\sim \{1,..,T\}} \sum\nolimits_i \text{CE}\left(r_{i},f_{\text{res}}(\vz_i^{(t)})\right)\end{matrix},
\end{align*}
with $f_{\text{noise}},f_{\text{res}}$ as denoising networks.
SiamDiff enhances this diffusion-based pre-training by generating correlated conformers via torsional perturbation and performing mutual denoising between two diffusion trajectories.

%% file: sections/04_experiment.tex
\section{Experiments}
\label{sec:exp}

\begin{table*}[t]
    \centering
    \caption{Evaluation results on EC, GO, PSR and MSP under various fusion schemes and structure encoders. 
    "PLM" and "Struct. Info." indicate the usage of protein language models and structural information in the model, respectively. 
    We employ underlining to highlight the best outcomes within each block and use bold symbols to highlight the best results for each task.
    }
    \label{tab:fusion}
    \begin{adjustbox}{max width=0.85\linewidth}
    \begin{threeparttable}
        \begin{tabular}{lccccccccccccc}
            \toprule
            \multirow{2}{*}{\bf{Method}} & \multirow{2}{*}{\bf{PLM}} & 
            \multirowcell{2}{\bf{Struct.}\\\bf{Info.}}
            &
            \multicolumn{1}{c}{\bf{EC}}&&
            \multicolumn{1}{c}{\bf{GO-BP}}&& \multicolumn{1}{c}{\bf{GO-MF}} & &\multicolumn{1}{c}{\bf{GO-CC}} && \bf{PSR} && \bf{MSP} \\
            \cmidrule{4-4}
            \cmidrule{6-6}
            \cmidrule{8-8}
            \cmidrule{10-10}
            \cmidrule{12-12}
            \cmidrule{14-14}
            & & & F\textsubscript{max} && F\textsubscript{max} && F\textsubscript{max} && F\textsubscript{max}  && Global $\rho$ && AUROC \\
            \midrule
            \bf{ProtBERT-BFD}\tnote{1} & \Checkmark & \XSolidBrush
             & 0.838 &&0.279 && 0.456 && 0.408 && - && -\\
            \bf{ESM-2-650M}\tnote{1} & \Checkmark & \XSolidBrush 
            & \underline{0.880} && \underline{0.460} && \underline{0.661} && \underline{0.445} && - && - \\
            \midrule
            \bf{\method} & \XSolidBrush & \Checkmark  & 0.730 && 0.356 && 0.503 && 0.414 && 0.708 && 0.549\\
            \bf{ESM-GearNet} & \\
            - w/ serial fusion & \multirow{3}{*}{\Checkmark} & \multirow{3}{*}{\Checkmark} & \bf{\underline{0.890}}	&& \bf{\underline{0.488}} && \bf{\underline{0.681}}  && \underline{0.464}  && \underline{0.829} && \bf{\underline{0.685}}\\
            - w/ parallel fusion & & & 0.792 && 0.384 && 0.573 && 0.407  &&0.760 && 0.644\\
            - w/ cross fusion &  & & 0.884 && 0.470 && 0.660  && 0.462 && 0.747 && 0.408\\
            \midrule
            \bf{GVP} & \XSolidBrush & \Checkmark 
            & 0.489 && 0.326 && 0.426 && 0.420 && 0.726 && \underline{0.664}\\
            \bf{ESM-GVP} & \\
            - w/ serial fusion & \multirow{3}{*}{\Checkmark} & \multirow{3}{*}{\Checkmark} & \underline{0.881} && \underline{0.473}&& \underline{0.668}  && \underline{0.485}  && \bf{\underline{0.866}} && 0.617\\
            - w/ parallel fusion & & & 0.872 && 0.446 && 0.657 && 0.455  && 0.702 && 0.592\\
            - w/ cross fusion & & & 0.880 && {0.465} && {0.664} && {0.469} && 0.764 && {0.583}\\
            \midrule
            \bf{CDConv} & \XSolidBrush & \Checkmark & 0.820 && 0.453 && 0.654  && \bf{\underline{0.479}}  && 0.786 && 0.529\\
            \bf{ESM-CDConv}\tnote{2} & \\
            - w/ serial fusion & \multirow{2}{*}{\Checkmark} & \multirow{2}{*}{\Checkmark} & \underline{0.880} && \underline{0.465}&& 0.658 && 0.475 && \underline{0.851} && {0.566}\\
            - w/ parallel fusion & & & 0.879 && 0.448 && \underline{0.662} && 0.455 && 0.803 && \underline{0.602}\\
            \bottomrule
        \end{tabular}
        \begin{tablenotes}
    \footnotesize
    \item[1] Protein language models do not take structures as input and thus cannot handle structure-related tasks like PSR and MSP.
    \item[2] Since CDConv does not yield residue-level representations, we cannot use cross fusion for ESM-CDConv.
    \end{tablenotes}
    \end{threeparttable}
    \end{adjustbox}
\end{table*}

\subsection{Setup}
\label{sec:setup}

In this section, we evaluate the effectiveness of our proposed methods on function annotation and structure property prediction tasks in Atom3D~\citep{townshend2021atom3d}.
Four key downstream tasks are considered:

    \textbf{1. Enzyme Commission (EC) Number Prediction:} This task involves predicting EC numbers that describe a protein's catalytic behavior in biochemical reactions. It's formulated as 538 binary classification problems based on the third and fourth levels of the EC tree~\citep{webb1992enzyme}. We use dataset splits from \citet{gligorijevic2021structure} and test on sequences with up to 95\% identity cutoff. 
    
    \textbf{2. Gene Ontology (GO) Term Prediction:} This benchmark includes three tasks: predicting a protein's biological process (BP), molecular function (MF), and cellular component (CC). Each task is framed as multiple binary classification problems based on GO term annotations. We employ dataset splits from \citet{gligorijevic2021structure} with a 95\% sequence identity cutoff. 
    
    \textbf{3. Protein Structure Ranking (PSR):} This task involves predicting global distance test scores for structure predictions submitted to the Critical Assessment of Structure Prediction (CASP)~\citep{kryshtafovych2019critical}. The dataset is partitioned by competition year. 
    
    \textbf{4. Mutation Stability Prediction (MSP):} The goal is to predict if a mutation enhances a protein complex's stability. The dataset is divided based on a 30\% sequence identity.

We employ the AlphaFold protein structure database v1~\citep{varadi2021alphafold} for pre-training, following~\citep{zhang2022protein}. This dataset encompasses 365K proteome-wide predictions from AlphaFold2.
For evaluation, we report the protein-centric maximum F-score (F\textsubscript{max}) for EC and GO prediction—common metrics in CAFA challenges~\citep{radivojac2013large}. Additionally, we present global Spearman correlation for PSR and AUROC for MSP.

\textbf{Training.}
Considering the model capacity and computational budget, we selected ESM-2-650M as the base PLM.
For structure encoders, we follow the original paper's default settings: 6 layers of GearNet with 512 hidden dimensions, 8 layers of CDConv with an initial hidden dimension of 256, and 5 layers of GVP with scalar features at 256 and vector features at 16 dimensions. We perform 50 epochs of pre-training on the AlphaFold Database, following the hyperparameters in~\citep{zhang2022protein}. Pre-training employs a batch size of 256 and a learning rate of 2e-4. For downstream evaluation, we utilize Adam optimizer with a batch size of 2 and a learning rate of 1e-4. These models are implemented using the TorchDrug library~\citep{zhu2022torchdrug} and trained across 4 A100 GPUs.

\begin{table*}[t]
    \centering
    \caption{Results of ESM-GearNet (serial fusion) pre-trained with six algorithms. The second and third column mark whether the pre-training methods include sequence and structure objectives, respectively. The best results are highlighted in bold. 
    }
    \label{tab:pre-train}
    \begin{adjustbox}{max width=0.9\linewidth}
        \begin{tabular}{lccccccccccccc}
            \toprule
            \multirow{2}{*}{\bf{Method}} & \multirowcell{2}{\bf{Sequence}\\\bf{Objective}} & \multirowcell{2}{\bf{Structure}\\\bf{Objective}} &
            \bf{EC}&&
            \bf{GO-BP}&& \bf{GO-MF} & &\bf{GO-CC} &&  \bf{PSR}&& \bf{MSP} \\
            \cmidrule{4-4}
            \cmidrule{6-6}
            \cmidrule{8-8}
            \cmidrule{10-10}
            \cmidrule{12-12}
            \cmidrule{14-14}
            & & & F\textsubscript{max} && F\textsubscript{max}  && F\textsubscript{max} && F\textsubscript{max} && Global $\rho$ && AUROC \\
            \midrule
            \bf{ESM-GearNet (serial fusion)} & - & - & 0.890 && 0.488 && 0.681 && 0.464  && 0.829 && 0.685\\
            - w/ Residue Type Prediction & \Checkmark & \XSolidBrush & 0.892 && {0.507} && 0.680 && 0.484 && 0.832 && 0.680\\
            - w/ Distance Prediction & \XSolidBrush & \Checkmark & 0.891 && 0.498 && 0.680 && 0.485 && {0.856} && 0.615\\
            - w/ Angle Prediction & \XSolidBrush & \Checkmark & 0.887 && 0.504 && 0.679 && 0.481 && {0.851} && \bf{0.702}\\
            - w/ Dihedral Prediction & \XSolidBrush & \Checkmark & 0.891 && 0.499 && 0.680 && 0.502 && 0.845 && 0.515\\
            - w/ Multiview Contrast & \XSolidBrush & \XSolidBrush  & {0.896} && \bf{0.514} && \bf{0.683} && {0.497} && 0.853 && 0.599\\
            - w/ SiamDiff & \Checkmark & \Checkmark & \bf{0.897} && 0.500 && {0.682} && \textbf{0.505} && \bf{0.863} && {0.692}\\
            \bottomrule
        \end{tabular}
    \end{adjustbox}
\end{table*}

\subsection{Results}

\paragraph{Evaluation of different fusion methods}
We evaluate our methods across the four tasks, employing the fusion of three structure encoders with ESM-2-650M using three distinct fusion strategies. 
The results are presented in Table~\ref{tab:fusion}. 
To provide context, we also include the outcomes of the structure encoders, along with two protein language models: ESM-2-650M and ProtBERT-BFD~\citep{elnaggar2020prottrans}.

First, upon intra-block result comparisons, it is evident that serial fusion, while simple in concept, is remarkably effective, surpassing the other two fusion strategies in the majority of tasks. 
The sole exceptions are ESM-CDConv on GO-MF and MSP, where the CDConv features yield marginal enhancements to PLMs for both fusion schemes.

Next, through inter-block result comparisons, we observe that while the raw performance of vanilla GearNet lags behind other encoders like CDConv, its integration with PLMs yields better outcomes, particularly for tasks like EC, GO-BP, and GO-MF. 
This underscores the efficacy of augmenting protein language representations onto structure encoders.
Notably, ESM-GVP's enhanced performance in PSR highlights the importance of model capcity in capturing structural details for such tasks.

Furthermore, upon comparing ESM-GearNet with ESM-2-650M, we observe substantial enhancements attributed to the incorporation of structural representations. 
This also enables it to effectively address structure-related tasks.

\paragraph{Effects of diminished learning rate}
\begin{figure}[t]
    \centering
    \includegraphics[width=\linewidth]{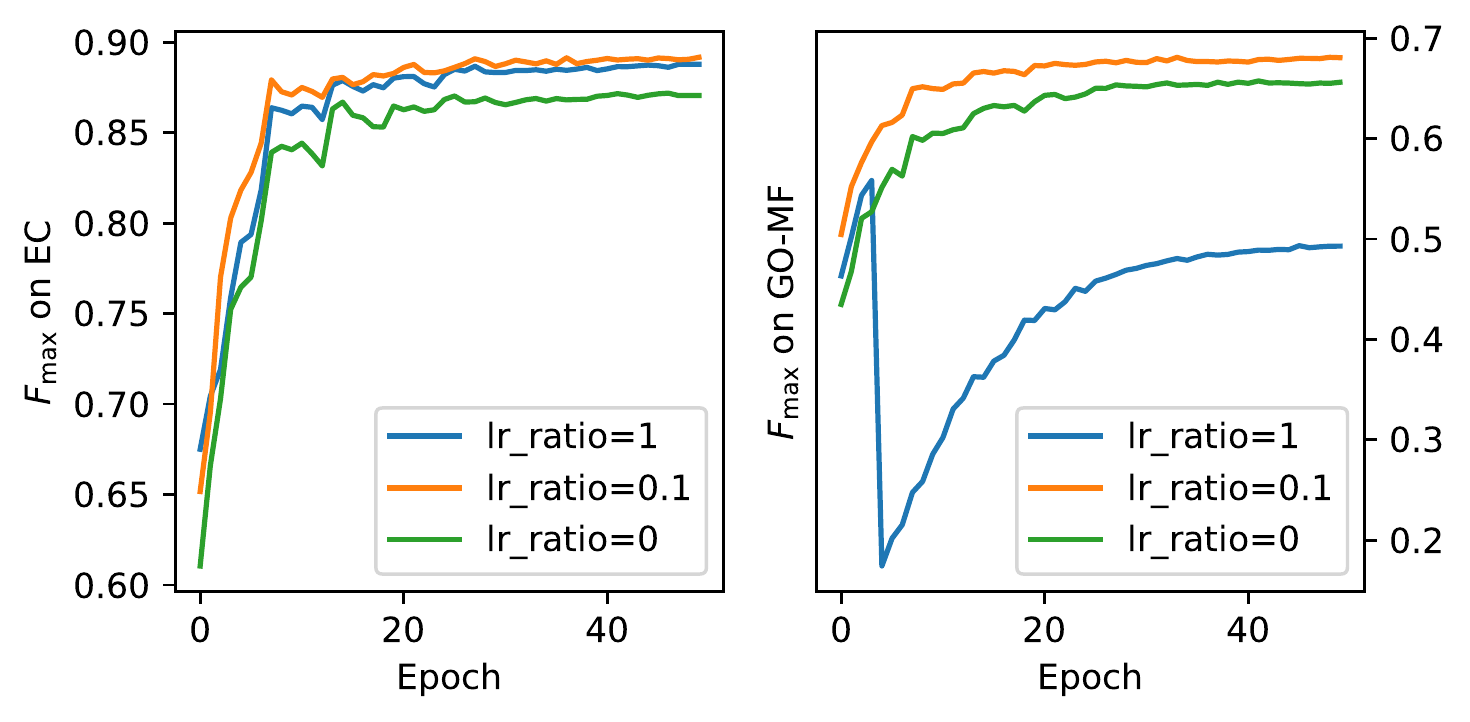}
    \vspace{-2em}
    \caption{ESM-GearNet (serial fusion) results on EC and GO-MF with different learning rate ratios.}
    \label{fig:lrratio}
    \vspace{-1em}
\end{figure}
To investigate the impact of reduced learning rates on representation fusion, we conducted experiments on EC and GO-MF using ESM-GearNet (serial fusion). We set different learning rate ratios for structure encoders relative to PLMs: 1, 0.1, and 0 (fixed). As illustrated in Figure~\ref{fig:lrratio}, the results consistently show that keeping PLMs fixed leads to inferior performance compared to fine-tuning them. When using equal learning rates for both PLMs and structure encoders, a notable performance drop occurs during GO-MF training, indicating significant deterioration of the PLM representations. This underscores the significance of employing reduced learning rates for PLMs to safeguard their representations from degradation.

\paragraph{Evaluation of different pre-training methods}
We choose the best-performing model, ESM-GearNet (serial fusion), and pre-train it with the introduced six algorithms.
The results are shown in Table~\ref{tab:pre-train}.
Notably, the top two pre-training methods are  Multiview Contrast and SiamDiff.
The former significantly enhances function annotation tasks such as EC, GO-BP, and GO-MF, while the latter excels in capturing structural intricacies for the GO-CC and PSR tasks. 
SiamDiff's superiority can be attributed to its incorporation of both sequential and structural pre-training objectives.
In contrast to methods that directly use either sequential or structural objectives, Multiview Contrast offers a more comprehensive consideration of sequence and structure dependencies.
By aligning representations from subsequences derived from the same protein, Multiview Contrast captures co-occurring sequences and structural motif dependencies. This utilization of ESM-2 and \method representations proves advantageous for function prediction.

\begin{table}[t]
    \centering
    \caption{Comparison between our methods with the state-of-the-art (SOTA) methods on benchmark tasks. 
    }
    \vspace{-0.5em}
    \label{tab:pre-train}
    \begin{adjustbox}{max width=\linewidth}
        \begin{tabular}{lccccccccccc}
            \toprule
            \multirow{2}{*}{\bf{Method}} & 
            \bf{EC}&&
            \bf{GO-BP}&& \bf{GO-MF} & &\bf{GO-CC}&& \bf{PSR} && \bf{MSP} \\
            \cmidrule{2-2}
            \cmidrule{4-4}
            \cmidrule{6-6}
            \cmidrule{8-8}
            \cmidrule{10-10}
            \cmidrule{12-12}
            & F\textsubscript{max} && F\textsubscript{max}  && F\textsubscript{max} && F\textsubscript{max} && Global $\rho$ && AUROC\\
            \midrule
            SOTA & 0.888 && 0.495 && 0.677 && \bf{0.551} && 0.862 && \bf{0.709}\\
            \midrule
            \bf{ESM-GearNet}  & 0.890	&& 0.488 && 0.681  && 0.464 && 0.829 && 0.685\\
            \small\bf{w/ pre-training}  & \bf{0.897} && \bf{0.514} && \bf{0.683} && 0.505 && \bf{0.863} && 0.702\\
            \bottomrule
        \end{tabular}
    \end{adjustbox}
    \vspace{-1em}
\end{table}

\paragraph{Comparison with state-of-the-art}
To showcase the robust performance of our proposed approaches, we compare them with previously established state-of-the-art methods, namely PromptProtein~\citep{wang2022multi} for EC and GO, and GVP~\citep{jing2021equivariant} for PSR and MSP tasks.
Our method demonstrates superior performance across most function annotation tasks, with the exception of GO-CC.
This outcome could be attributed to the fact that GO-CC pertains to predicting the cellular component of a protein's function, which may be less directly related to the protein's primary function. Additionally, our approach yields competitive results in the PSR and MSP tasks, aligning with the previous state-of-the-art performance.

%% file: sections/05_conclusion.tex
\section{Conclusions}
\label{sec:conclusion}

This study presents a comprehensive exploration of joint protein representation learning, effectively fusing protein language models (PLMs) and structure encoders to harness the strengths of both domains. The integration of ESM-2 with diverse structure encoders, alongside the introduction of innovative fusion strategies, has yielded valuable insights into effective joint representation learning. Our findings highlight the mutually beneficial relationship between sequence and structure information during pre-training, emphasizing the importance of a holistic approach. By achieving new state-of-the-art results in tasks such as Enzyme Commission number and Gene Ontology term annotation, our work not only advances the adaptation of PLMs and structure encoders but also holds broader implications for protein representation learning. 

%% file: main.bbl
\begin{thebibliography}{57}
\providecommand{\natexlab}[1]{#1}

\bibitem[{Alley et~al.(2019)Alley, Khimulya, Biswas, AlQuraishi, and
  Church}]{alley2019unified}
Alley, E.~C.; Khimulya, G.; Biswas, S.; AlQuraishi, M.; and Church, G.~M. 2019.
\newblock Unified rational protein engineering with sequence-based deep
  representation learning.
\newblock \emph{Nature methods}, 16(12): 1315--1322.

\bibitem[{Baek et~al.(2021)Baek, DiMaio, Anishchenko, Dauparas, Ovchinnikov,
  Lee, Wang, Cong, Kinch, Schaeffer et~al.}]{baek2021accurate}
Baek, M.; DiMaio, F.; Anishchenko, I.; Dauparas, J.; Ovchinnikov, S.; Lee,
  G.~R.; Wang, J.; Cong, Q.; Kinch, L.~N.; Schaeffer, R.~D.; et~al. 2021.
\newblock Accurate prediction of protein structures and interactions using a
  three-track neural network.
\newblock \emph{Science}, 373(6557): 871--876.

\bibitem[{Berman et~al.(2000)Berman, Westbrook, Feng, Gilliland, Bhat, Weissig,
  Shindyalov, and Bourne}]{berman2000protein}
Berman, H.~M.; Westbrook, J.; Feng, Z.; Gilliland, G.; Bhat, T.~N.; Weissig,
  H.; Shindyalov, I.~N.; and Bourne, P.~E. 2000.
\newblock The protein data bank.
\newblock \emph{Nucleic acids research}, 28(1): 235--242.

\bibitem[{Brown et~al.(2020)Brown, Mann, Ryder, Subbiah, Kaplan, Dhariwal,
  Neelakantan, Shyam, Sastry, Askell, Agarwal, Herbert-Voss, Krueger, Henighan,
  Child, Ramesh, Ziegler, Wu, Winter, Hesse, Chen, Sigler, Litwin, Gray, Chess,
  Clark, Berner, McCandlish, Radford, Sutskever, and
  Amodei}]{NEURIPS2020_1457c0d6}
Brown, T.; Mann, B.; Ryder, N.; Subbiah, M.; Kaplan, J.~D.; Dhariwal, P.;
  Neelakantan, A.; Shyam, P.; Sastry, G.; Askell, A.; Agarwal, S.;
  Herbert-Voss, A.; Krueger, G.; Henighan, T.; Child, R.; Ramesh, A.; Ziegler,
  D.; Wu, J.; Winter, C.; Hesse, C.; Chen, M.; Sigler, E.; Litwin, M.; Gray,
  S.; Chess, B.; Clark, J.; Berner, C.; McCandlish, S.; Radford, A.; Sutskever,
  I.; and Amodei, D. 2020.
\newblock In \emph{Advances in Neural Information Processing Systems}.

\bibitem[{Chen et~al.(2023{\natexlab{a}})Chen, Cheng, Geng, Li, Zeng, Wang,
  Gong, Liu, Zeng, Dong et~al.}]{chen2023xtrimopglm}
Chen, B.; Cheng, X.; Geng, Y.-a.; Li, S.; Zeng, X.; Wang, B.; Gong, J.; Liu,
  C.; Zeng, A.; Dong, Y.; et~al. 2023{\natexlab{a}}.
\newblock xtrimopglm: Unified 100b-scale pre-trained transformer for
  deciphering the language of protein.
\newblock \emph{bioRxiv}, 2023--07.

\bibitem[{Chen et~al.(2023{\natexlab{b}})Chen, Zhou, Wang, Liu, and
  Dou}]{chen2022structure}
Chen, C.~S.; Zhou, J.; Wang, F.; Liu, X.; and Dou, D. 2023{\natexlab{b}}.
\newblock {Structure-aware protein self-supervised learning}.
\newblock \emph{Bioinformatics}, 39(4): btad189.

\bibitem[{Chen et~al.(2020)Chen, Kornblith, Norouzi, and
  Hinton}]{chen2020simple}
Chen, T.; Kornblith, S.; Norouzi, M.; and Hinton, G. 2020.
\newblock A simple framework for contrastive learning of visual
  representations.
\newblock In \emph{International conference on machine learning}, 1597--1607.
  PMLR.

\bibitem[{Corso et~al.(2023)Corso, Stärk, Jing, Barzilay, and
  Jaakkola}]{corso2023diffdock}
Corso, G.; Stärk, H.; Jing, B.; Barzilay, R.; and Jaakkola, T. 2023.
\newblock DiffDock: Diffusion Steps, Twists, and Turns for Molecular Docking.
\newblock \emph{International Conference on Learning Representations (ICLR)}.

\bibitem[{Dauparas et~al.(2022)Dauparas, Anishchenko, Bennett, Bai, Ragotte,
  Milles, Wicky, Courbet, de~Haas, Bethel et~al.}]{dauparas2022robust}
Dauparas, J.; Anishchenko, I.; Bennett, N.; Bai, H.; Ragotte, R.~J.; Milles,
  L.~F.; Wicky, B.~I.; Courbet, A.; de~Haas, R.~J.; Bethel, N.; et~al. 2022.
\newblock Robust deep learning--based protein sequence design using
  ProteinMPNN.
\newblock \emph{Science}, 378(6615): 49--56.

\bibitem[{Devlin et~al.(2019)Devlin, Chang, Lee, and
  Toutanova}]{devlin2018bert}
Devlin, J.; Chang, M.-W.; Lee, K.; and Toutanova, K. 2019.
\newblock {BERT}: Pre-training of Deep Bidirectional Transformers for Language
  Understanding.
\newblock In \emph{Proceedings of the 2019 Conference of the North {A}merican
  Chapter of the Association for Computational Linguistics: Human Language
  Technologies, Volume 1 (Long and Short Papers)}, 4171--4186. Minneapolis,
  Minnesota: Association for Computational Linguistics.

\bibitem[{Elnaggar et~al.(2023{\natexlab{a}})Elnaggar, Essam, Salah-Eldin,
  Moustafa, Elkerdawy, Rochereau, and Rost}]{elnaggar2023ankh}
Elnaggar, A.; Essam, H.; Salah-Eldin, W.; Moustafa, W.; Elkerdawy, M.;
  Rochereau, C.; and Rost, B. 2023{\natexlab{a}}.
\newblock Ankh: Optimized Protein Language Model Unlocks General-Purpose
  Modelling.
\newblock \emph{bioRxiv}, 2023--01.

\bibitem[{Elnaggar et~al.(2023{\natexlab{b}})Elnaggar, Essam, Salah-Eldin,
  Moustafa, Elkerdawy, Rochereau, and Rost}]{Elnaggar2023AnkhO}
Elnaggar, A.; Essam, H.; Salah-Eldin, W.; Moustafa, W. F.~O.; Elkerdawy, M.;
  Rochereau, C.; and Rost, B. 2023{\natexlab{b}}.
\newblock Ankh: Optimized Protein Language Model Unlocks General-Purpose
  Modelling.
\newblock \emph{bioRxiv}.

\bibitem[{Elnaggar et~al.(2021{\natexlab{a}})Elnaggar, Heinzinger, Dallago,
  Rehawi, Wang, Jones, Gibbs, Feher, Angerer, Steinegger
  et~al.}]{elnaggar2021prottrans}
Elnaggar, A.; Heinzinger, M.; Dallago, C.; Rehawi, G.; Wang, Y.; Jones, L.;
  Gibbs, T.; Feher, T.; Angerer, C.; Steinegger, M.; et~al. 2021{\natexlab{a}}.
\newblock Prottrans: Toward understanding the language of life through
  self-supervised learning.
\newblock \emph{IEEE transactions on pattern analysis and machine
  intelligence}, 44(10): 7112--7127.

\bibitem[{Elnaggar et~al.(2021{\natexlab{b}})Elnaggar, Heinzinger, Dallago,
  Rehawi, Yu, Jones, Gibbs, Feher, Angerer, Steinegger, Bhowmik, and
  Rost}]{elnaggar2020prottrans}
Elnaggar, A.; Heinzinger, M.; Dallago, C.; Rehawi, G.; Yu, W.; Jones, L.;
  Gibbs, T.; Feher, T.; Angerer, C.; Steinegger, M.; Bhowmik, D.; and Rost, B.
  2021{\natexlab{b}}.
\newblock ProtTrans: Towards Cracking the Language of Lifes Code Through
  Self-Supervised Deep Learning and High Performance Computing.
\newblock \emph{IEEE Transactions on Pattern Analysis and Machine
  Intelligence}, 1--1.

\bibitem[{Fan et~al.(2023)Fan, Wang, Yang, and
  Kankanhalli}]{fan2023continuousdiscrete}
Fan, H.; Wang, Z.; Yang, Y.; and Kankanhalli, M. 2023.
\newblock Continuous-Discrete Convolution for Geometry-Sequence Modeling in
  Proteins.
\newblock In \emph{The Eleventh International Conference on Learning
  Representations}.

\bibitem[{Gainza et~al.(2020)Gainza, Sverrisson, Monti, Rodola, Boscaini,
  Bronstein, and Correia}]{gainza2020deciphering}
Gainza, P.; Sverrisson, F.; Monti, F.; Rodola, E.; Boscaini, D.; Bronstein, M.;
  and Correia, B. 2020.
\newblock Deciphering interaction fingerprints from protein molecular surfaces
  using geometric deep learning.
\newblock \emph{Nature Methods}, 17(2): 184--192.

\bibitem[{Gao, Tan, and Li(2023)}]{gao2023pifold}
Gao, Z.; Tan, C.; and Li, S.~Z. 2023.
\newblock PiFold: Toward effective and efficient protein inverse folding.
\newblock In \emph{The Eleventh International Conference on Learning
  Representations}.

\bibitem[{Gligorijevi{\'c} et~al.(2021)Gligorijevi{\'c}, Renfrew, Kosciolek,
  Leman, Berenberg, Vatanen, Chandler, Taylor, Fisk, Vlamakis
  et~al.}]{gligorijevic2021structure}
Gligorijevi{\'c}, V.; Renfrew, P.~D.; Kosciolek, T.; Leman, J.~K.; Berenberg,
  D.; Vatanen, T.; Chandler, C.; Taylor, B.~C.; Fisk, I.~M.; Vlamakis, H.;
  et~al. 2021.
\newblock Structure-based protein function prediction using graph convolutional
  networks.
\newblock \emph{Nature communications}, 12(1): 1--14.

\bibitem[{Guo et~al.(2022)Guo, Wu, Ma, and Huang}]{Guo2022SelfSupervisedPF}
Guo, Y.; Wu, J.; Ma, H.; and Huang, J. 2022.
\newblock Self-Supervised Pre-training for Protein Embeddings Using Tertiary
  Structures.
\newblock In \emph{AAAI}.

\bibitem[{He et~al.(2021)He, Zhang, Wu, Xia, Ju, Zhang, Liu, Xia, Zhu, Deng
  et~al.}]{he2021pre}
He, L.; Zhang, S.; Wu, L.; Xia, H.; Ju, F.; Zhang, H.; Liu, S.; Xia, Y.; Zhu,
  J.; Deng, P.; et~al. 2021.
\newblock Pre-training Co-evolutionary Protein Representation via A Pairwise
  Masked Language Model.
\newblock \emph{arXiv preprint arXiv:2110.15527}.

\bibitem[{Heinzinger et~al.(2023)Heinzinger, Weissenow, Sanchez, Henkel,
  Steinegger, and Rost}]{heinzinger2023prostt5}
Heinzinger, M.; Weissenow, K.; Sanchez, J.~G.; Henkel, A.; Steinegger, M.; and
  Rost, B. 2023.
\newblock ProstT5: Bilingual Language Model for Protein Sequence and Structure.
\newblock \emph{bioRxiv}, 2023--07.

\bibitem[{Hermosilla and Ropinski(2022)}]{hermosilla2022contrastive}
Hermosilla, P.; and Ropinski, T. 2022.
\newblock Contrastive representation learning for 3d protein structures.
\newblock \emph{arXiv preprint arXiv:2205.15675}.

\bibitem[{Hermosilla et~al.(2021)Hermosilla, Sch{\"a}fer, Lang, Fackelmann,
  V{\'a}zquez, Kozl{\'\i}kov{\'a}, Krone, Ritschel, and
  Ropinski}]{hermosilla2020intrinsic}
Hermosilla, P.; Sch{\"a}fer, M.; Lang, M.; Fackelmann, G.; V{\'a}zquez, P.~P.;
  Kozl{\'\i}kov{\'a}, B.; Krone, M.; Ritschel, T.; and Ropinski, T. 2021.
\newblock Intrinsic-Extrinsic Convolution and Pooling for Learning on 3D
  Protein Structures.
\newblock \emph{International Conference on Learning Representations}.

\bibitem[{Hesslow et~al.(2022)Hesslow, Zanichelli, Notin, Poli, and
  Marks}]{hesslow2022rita}
Hesslow, D.; Zanichelli, N.; Notin, P.; Poli, I.; and Marks, D. 2022.
\newblock Rita: a study on scaling up generative protein sequence models.
\newblock In \emph{2022 ICML Workshop on Computational Biology}.

\bibitem[{Hsu et~al.(2022)Hsu, Verkuil, Liu, Lin, Hie, Sercu, Lerer, and
  Rives}]{hsu2022learning}
Hsu, C.; Verkuil, R.; Liu, J.; Lin, Z.; Hie, B.; Sercu, T.; Lerer, A.; and
  Rives, A. 2022.
\newblock Learning inverse folding from millions of predicted structures.
\newblock \emph{ICML}.

\bibitem[{Huang et~al.(2023)Huang, Wu, Lin, Zheng, Wang, and
  Li}]{huang2023data}
Huang, Y.; Wu, L.; Lin, H.; Zheng, J.; Wang, G.; and Li, S.~Z. 2023.
\newblock Data-Efficient Protein 3D Geometric Pretraining via Refinement of
  Diffused Protein Structure Decoy.
\newblock \emph{arXiv preprint arXiv:2302.10888}.

\bibitem[{Jing et~al.(2021)Jing, Eismann, Soni, and Dror}]{jing2021equivariant}
Jing, B.; Eismann, S.; Soni, P.~N.; and Dror, R.~O. 2021.
\newblock Learning from Protein Structure with Geometric Vector Perceptrons.
\newblock In \emph{International Conference on Learning Representations}.

\bibitem[{Jumper et~al.(2021)Jumper, Evans, Pritzel, Green, Figurnov,
  Ronneberger, Tunyasuvunakool, Bates, {\v{Z}}{\'\i}dek, Potapenko
  et~al.}]{jumper2021highly}
Jumper, J.; Evans, R.; Pritzel, A.; Green, T.; Figurnov, M.; Ronneberger, O.;
  Tunyasuvunakool, K.; Bates, R.; {\v{Z}}{\'\i}dek, A.; Potapenko, A.; et~al.
  2021.
\newblock Highly accurate protein structure prediction with AlphaFold.
\newblock \emph{Nature}, 596(7873): 583--589.

\bibitem[{Kryshtafovych et~al.(2019)Kryshtafovych, Schwede, Topf, Fidelis, and
  Moult}]{kryshtafovych2019critical}
Kryshtafovych, A.; Schwede, T.; Topf, M.; Fidelis, K.; and Moult, J. 2019.
\newblock Critical assessment of methods of protein structure prediction
  (CASP)—Round XIII.
\newblock \emph{Proteins: Structure, Function, and Bioinformatics}, 87(12):
  1011--1020.

\bibitem[{Lin et~al.(2023)Lin, Akin, Rao, Hie, Zhu, Lu, Smetanin, Verkuil,
  Kabeli, Shmueli et~al.}]{lin2023evolutionary}
Lin, Z.; Akin, H.; Rao, R.; Hie, B.; Zhu, Z.; Lu, W.; Smetanin, N.; Verkuil,
  R.; Kabeli, O.; Shmueli, Y.; et~al. 2023.
\newblock Evolutionary-scale prediction of atomic-level protein structure with
  a language model.
\newblock \emph{Science}, 379(6637): 1123--1130.

\bibitem[{Lu et~al.(2020)Lu, Zhang, Ghassemi, and Moses}]{lu2020self}
Lu, A.~X.; Zhang, H.; Ghassemi, M.; and Moses, A.~M. 2020.
\newblock Self-supervised contrastive learning of protein representations by
  mutual information maximization.
\newblock \emph{BioRxiv}.

\bibitem[{Madani et~al.(2023)Madani, Krause, Greene, Subramanian, Mohr, Holton,
  Olmos~Jr, Xiong, Sun, Socher et~al.}]{madani2023large}
Madani, A.; Krause, B.; Greene, E.~R.; Subramanian, S.; Mohr, B.~P.; Holton,
  J.~M.; Olmos~Jr, J.~L.; Xiong, C.; Sun, Z.~Z.; Socher, R.; et~al. 2023.
\newblock Large language models generate functional protein sequences across
  diverse families.
\newblock \emph{Nature Biotechnology}, 1--8.

\bibitem[{Meier et~al.(2021)Meier, Rao, Verkuil, Liu, Sercu, and
  Rives}]{meier2021language}
Meier, J.; Rao, R.; Verkuil, R.; Liu, J.; Sercu, T.; and Rives, A. 2021.
\newblock Language models enable zero-shot prediction of the effects of
  mutations on protein function.
\newblock In Beygelzimer, A.; Dauphin, Y.; Liang, P.; and Vaughan, J.~W., eds.,
  \emph{Advances in Neural Information Processing Systems}.

\bibitem[{Notin et~al.(2022)Notin, Dias, Frazer, Hurtado, Gomez, Marks, and
  Gal}]{notin2022tranception}
Notin, P.; Dias, M.; Frazer, J.; Hurtado, J.~M.; Gomez, A.~N.; Marks, D.; and
  Gal, Y. 2022.
\newblock Tranception: protein fitness prediction with autoregressive
  transformers and inference-time retrieval.
\newblock In \emph{International Conference on Machine Learning}, 16990--17017.
  PMLR.

\bibitem[{Ponting and Russell(2002)}]{Ponting2002TheNH}
Ponting, C.~P.; and Russell, R.~R. 2002.
\newblock The natural history of protein domains.
\newblock \emph{Annual review of biophysics and biomolecular structure}, 31:
  45--71.

\bibitem[{Radivojac et~al.(2013)Radivojac, Clark, Oron, Schnoes, Wittkop,
  Sokolov, Graim, Funk, Verspoor, Ben-Hur et~al.}]{radivojac2013large}
Radivojac, P.; Clark, W.~T.; Oron, T.~R.; Schnoes, A.~M.; Wittkop, T.; Sokolov,
  A.; Graim, K.; Funk, C.; Verspoor, K.; Ben-Hur, A.; et~al. 2013.
\newblock A large-scale evaluation of computational protein function
  prediction.
\newblock \emph{Nature methods}, 10(3): 221--227.

\bibitem[{Rao et~al.(2019)Rao, Bhattacharya, Thomas, Duan, Chen, Canny, Abbeel,
  and Song}]{tape2019}
Rao, R.; Bhattacharya, N.; Thomas, N.; Duan, Y.; Chen, X.; Canny, J.; Abbeel,
  P.; and Song, Y.~S. 2019.
\newblock Evaluating Protein Transfer Learning with TAPE.
\newblock In \emph{Advances in Neural Information Processing Systems}.

\bibitem[{Rives et~al.(2021)Rives, Meier, Sercu, Goyal, Lin, Liu, Guo, Ott,
  Zitnick, Ma et~al.}]{rives2021biological}
Rives, A.; Meier, J.; Sercu, T.; Goyal, S.; Lin, Z.; Liu, J.; Guo, D.; Ott, M.;
  Zitnick, C.~L.; Ma, J.; et~al. 2021.
\newblock Biological structure and function emerge from scaling unsupervised
  learning to 250 million protein sequences.
\newblock \emph{Proceedings of the National Academy of Sciences}, 118(15).

\bibitem[{Shanehsazzadeh, Belanger, and
  Dohan(2020)}]{shanehsazzadeh2020transfer}
Shanehsazzadeh, A.; Belanger, D.; and Dohan, D. 2020.
\newblock Is transfer learning necessary for protein landscape prediction?
\newblock \emph{arXiv preprint arXiv:2011.03443}.

\bibitem[{Sverrisson et~al.(2021)Sverrisson, Feydy, Correia, and
  Bronstein}]{sverrisson2021fast}
Sverrisson, F.; Feydy, J.; Correia, B.~E.; and Bronstein, M.~M. 2021.
\newblock Fast end-to-end learning on protein surfaces.
\newblock In \emph{Proceedings of the IEEE/CVF Conference on Computer Vision
  and Pattern Recognition}, 15272--15281.

\bibitem[{Townshend et~al.(2021)Townshend, V{\"o}gele, Suriana, Derry, Powers,
  Laloudakis, Balachandar, Jing, Anderson, Eismann, Kondor, Altman, and
  Dror}]{townshend2021atom3d}
Townshend, R. J.~L.; V{\"o}gele, M.; Suriana, P.~A.; Derry, A.; Powers, A.;
  Laloudakis, Y.; Balachandar, S.; Jing, B.; Anderson, B.~M.; Eismann, S.;
  Kondor, R.; Altman, R.; and Dror, R.~O. 2021.
\newblock {ATOM}3D: Tasks on Molecules in Three Dimensions.
\newblock In \emph{Thirty-fifth Conference on Neural Information Processing
  Systems Datasets and Benchmarks Track (Round 1)}.

\bibitem[{Trippe et~al.(2023)Trippe, Yim, Tischer, Baker, Broderick, Barzilay,
  and Jaakkola}]{trippe2023diffusion}
Trippe, B.~L.; Yim, J.; Tischer, D.; Baker, D.; Broderick, T.; Barzilay, R.;
  and Jaakkola, T.~S. 2023.
\newblock Diffusion Probabilistic Modeling of Protein Backbones in 3D for the
  motif-scaffolding problem.
\newblock In \emph{The Eleventh International Conference on Learning
  Representations}.

\bibitem[{Varadi et~al.(2021)Varadi, Anyango, Deshpande, Nair, Natassia,
  Yordanova, Yuan, Stroe, Wood, Laydon et~al.}]{varadi2021alphafold}
Varadi, M.; Anyango, S.; Deshpande, M.; Nair, S.; Natassia, C.; Yordanova, G.;
  Yuan, D.; Stroe, O.; Wood, G.; Laydon, A.; et~al. 2021.
\newblock AlphaFold Protein Structure Database: massively expanding the
  structural coverage of protein-sequence space with high-accuracy models.
\newblock \emph{Nucleic acids research}.

\bibitem[{Vaswani et~al.(2017)Vaswani, Shazeer, Parmar, Uszkoreit, Jones,
  Gomez, Kaiser, and Polosukhin}]{vaswani2017attention}
Vaswani, A.; Shazeer, N.; Parmar, N.; Uszkoreit, J.; Jones, L.; Gomez, A.~N.;
  Kaiser, {\L}.; and Polosukhin, I. 2017.
\newblock Attention is all you need.
\newblock In \emph{Advances in neural information processing systems},
  5998--6008.

\bibitem[{Wang et~al.(2022{\natexlab{a}})Wang, Combs, Brand, Calvo, Xu, Price,
  Golovach, Salawu, Wise, Ponnapalli et~al.}]{wang2022lm}
Wang, Z.; Combs, S.~A.; Brand, R.; Calvo, M.~R.; Xu, P.; Price, G.; Golovach,
  N.; Salawu, E.~O.; Wise, C.~J.; Ponnapalli, S.~P.; et~al. 2022{\natexlab{a}}.
\newblock Lm-gvp: an extensible sequence and structure informed deep learning
  framework for protein property prediction.
\newblock \emph{Scientific reports}, 12(1): 6832.

\bibitem[{Wang et~al.(2022{\natexlab{b}})Wang, Zhang, Shuang-Wei, Yu, Jin,
  Gong, and Chen}]{wang2022multi}
Wang, Z.; Zhang, Q.; Shuang-Wei, H.; Yu, H.; Jin, X.; Gong, Z.; and Chen, H.
  2022{\natexlab{b}}.
\newblock Multi-level Protein Structure Pre-training via Prompt Learning.
\newblock In \emph{The Eleventh International Conference on Learning
  Representations}.

\bibitem[{Webb et~al.(1992)Webb, Phelps, Bienkowski, Digrazia, White, and
  Sayler}]{webb1992enzyme}
Webb, O.~F.; Phelps, T.~J.; Bienkowski, P.~R.; Digrazia, P.~M.; White, D.~C.;
  and Sayler, G.~S. 1992.
\newblock Enzyme nomenclature.

\bibitem[{Wu et~al.(2022{\natexlab{a}})Wu, Zhang, Radev, Wang, Jin, Jiang, Niu,
  and Li}]{wu2022pre}
Wu, F.; Zhang, Q.; Radev, D.; Wang, Y.; Jin, X.; Jiang, Y.; Niu, Z.; and Li,
  S.~Z. 2022{\natexlab{a}}.
\newblock Pre-training of Deep Protein Models with Molecular Dynamics
  Simulations for Drug Binding.
\newblock \emph{arXiv preprint arXiv:2204.08663}.

\bibitem[{Wu et~al.(2022{\natexlab{b}})Wu, Yang, Berg, Zou, Lu, and
  Amini}]{wu2022protein}
Wu, K.~E.; Yang, K.~K.; Berg, R. v.~d.; Zou, J.~Y.; Lu, A.~X.; and Amini, A.~P.
  2022{\natexlab{b}}.
\newblock Protein structure generation via folding diffusion.
\newblock \emph{arXiv preprint arXiv:2209.15611}.

\bibitem[{Xu et~al.(2023{\natexlab{a}})Xu, Guo, Xu, Tang, Chen, and
  Tian}]{xu2023eurnet}
Xu, M.; Guo, Y.; Xu, Y.; Tang, J.; Chen, X.; and Tian, Y. 2023{\natexlab{a}}.
\newblock EurNet: Efficient Multi-Range Relational Modeling of Protein
  Structure.
\newblock In \emph{ICLR 2023 - Machine Learning for Drug Discovery workshop}.

\bibitem[{Xu et~al.(2023{\natexlab{b}})Xu, Yuan, Miret, and
  Tang}]{xu2023protst}
Xu, M.; Yuan, X.; Miret, S.; and Tang, J. 2023{\natexlab{b}}.
\newblock {P}rot{ST}: Multi-Modality Learning of Protein Sequences and
  Biomedical Texts.
\newblock In Krause, A.; Brunskill, E.; Cho, K.; Engelhardt, B.; Sabato, S.;
  and Scarlett, J., eds., \emph{Proceedings of the 40th International
  Conference on Machine Learning}, volume 202 of \emph{Proceedings of Machine
  Learning Research}, 38749--38767. PMLR.

\bibitem[{Xu et~al.(2022)Xu, Zhang, Lu, Zhu, Zhang, Ma, Liu, and
  Tang}]{xu2022peer}
Xu, M.; Zhang, Z.; Lu, J.; Zhu, Z.; Zhang, Y.; Ma, C.; Liu, R.; and Tang, J.
  2022.
\newblock {PEER}: A Comprehensive and Multi-Task Benchmark for Protein Sequence
  Understanding.
\newblock In \emph{Thirty-sixth Conference on Neural Information Processing
  Systems Datasets and Benchmarks Track}.

\bibitem[{Zhang et~al.(2022{\natexlab{a}})Zhang, Bi, Liang, Cheng, Hong, Deng,
  Zhang, Lian, and Chen}]{zhang2022ontoprotein}
Zhang, N.; Bi, Z.; Liang, X.; Cheng, S.; Hong, H.; Deng, S.; Zhang, Q.; Lian,
  J.; and Chen, H. 2022{\natexlab{a}}.
\newblock OntoProtein: Protein Pretraining With Gene Ontology Embedding.
\newblock In \emph{International Conference on Learning Representations}.

\bibitem[{Zhang et~al.(2023{\natexlab{a}})Zhang, Cai, Shi, and
  Tang}]{zhang2023ebind}
Zhang, Y.; Cai, H.; Shi, C.; and Tang, J. 2023{\natexlab{a}}.
\newblock E3Bind: An End-to-End Equivariant Network for Protein-Ligand Docking.
\newblock In \emph{The Eleventh International Conference on Learning
  Representations}.

\bibitem[{Zhang et~al.(2022{\natexlab{b}})Zhang, Xu, Jamasb, Chenthamarakshan,
  Lozano, Das, and Tang}]{zhang2022protein}
Zhang, Z.; Xu, M.; Jamasb, A.~R.; Chenthamarakshan, V.; Lozano, A.; Das, P.;
  and Tang, J. 2022{\natexlab{b}}.
\newblock Protein Representation Learning by Geometric Structure Pretraining.
\newblock In \emph{First Workshop on Pre-training: Perspectives, Pitfalls, and
  Paths Forward at ICML 2022}.

\bibitem[{Zhang et~al.(2023{\natexlab{b}})Zhang, Xu, Lozano, Chenthamarakshan,
  Das, and Tang}]{zhang2023siamdiff}
Zhang, Z.; Xu, M.; Lozano, A.; Chenthamarakshan, V.; Das, P.; and Tang, J.
  2023{\natexlab{b}}.
\newblock Pre-Training Protein Encoder via Siamese Sequence-Structure Diffusion
  Trajectory Prediction.
\newblock In \emph{Advances in Neural Information Processing Systems}.

\bibitem[{Zhu et~al.(2022)Zhu, Shi, Zhang, Liu, Xu, Yuan, Zhang, Chen, Cai, Lu
  et~al.}]{zhu2022torchdrug}
Zhu, Z.; Shi, C.; Zhang, Z.; Liu, S.; Xu, M.; Yuan, X.; Zhang, Y.; Chen, J.;
  Cai, H.; Lu, J.; et~al. 2022.
\newblock TorchDrug: A Powerful and Flexible Machine Learning Platform for Drug
  Discovery.
\newblock \emph{arXiv preprint arXiv:2202.08320}.

\end{thebibliography}
